\documentstyle[preprint,pra,floats,psfig,aps]{revtex}

\title{Hyperfine spectra of CH$_3$F nuclear spin conversion}
\author{Pavel~L.~Chapovsky}
\address{Institute of Automation and Electrometry,
          Russian Academy of Sciences, 630090 Novosibirsk, Russia; and 
          Laboratoire de Physique des Lasers, Atomes et Molecules,\\ 
Universit\'e des Sciences et Technologies de Lille, 
F-59655 Villeneuve d'Ascq Cedex, France}          
         
\date{\today}

\begin{document}
\draft
\maketitle

\begin{abstract}             
A theoretical model of hyperfine spectra of nuclear spin conversion
of molecular isomers has been developed.  The model takes into account the 
nuclear spin-spin and spin-rotation interactions, as well as the
saturation of intramolecular mixing of molecular 
ortho and para  states. The model has been applied to 
hyperfine spectra of nuclear spin conversion in $^{13}$CH$_3$F 
molecules subjected to an external electric field. Conditions under which 
the hyperfine structure in the spectra can be resolved have been determined.
\end{abstract}
\vspace{1cm}
\pacs{34.30.+h, 35.20.Sd, 33.50.-j\\
      Key words: Molecular spin isomers, conversion, hyperfine spectra.}

\section{Introduction}

Recently, much attention has been devoted to the investigation of the 
level-crossing resonances in the spin isomer conversion 
({\it conversion spectra}) in CH$_3$F 
\cite{Nagels95CPL,Chap97,Bahloul98JPB,Bahloul98,Nagels96PRL,Nagels98,Cosleou00EPJD}.
The goal was to test the mechanism of the ortho-para conversion in
this molecule. The importance of these efforts is clear. CH$_3$F is
the first and until now the only polyatomic molecule in which a  
spin conversion mechanism was identified. On the other hand, one may
expect a similar conversion mechanism in other symmetrical molecules too. 

The essence of the isomer conversion spectra 
consists of the following. An external homogeneous electric field
splits the molecular states. At some electric field the ortho and para states
of the molecule cross which, speeds up the rate of conversion.
Scanning of the electric field gives the field dependence of conversion rate 
which is the conversion spectrum. Measurements of the conversion spectra  
are presently performed at room temperature. 

The usefulness of the conversion spectra should not 
be limited to the investigation of the conversion
itself. In fact, this is a new physical effect which can find various 
applications. For example, it was proposed \cite{Bahloul98JPB}
to measure with the help of the conversion spectra 
characteristics of molecular spin-rotation 
interaction which are hardly accessible by ordinary  methods.
Many of these applications require a deeper understanding of the phenomenon.
Up to now, only the rotational structure in the conversion spectra  
resulting from $M$-splitting of molecular states has been considered. 
This is sufficient for the conversion spectra of  
$\sim$1~MHz resolution. Higher resolution needs an account of the 
hyperfine structures in the conversion spectra, which
has not been done before. 
In this paper we perform such an analysis and determine the 
conditions at which these hyperfine structures can be observed. 
We check also an accuracy of the simplified approach to the 
conversion spectra in 
which hyperfine splitting of molecular states is neglected.
The calculations are done for $^{13}$CH$_3$F because for this 
molecule the rotational structure in the conversion spectra was experimentally 
observed recently \cite{Nagels96PRL}. The spectral resolution achieved
already in \cite{Nagels96PRL} was better than 6~MHz. It makes the $^{13}$CH$_3$F
molecule the best candidate for the observation of the hyperfine 
structures in the conversion spectra.

\section{Quantum relaxation of spin isomers}

Ordinary gaseous relaxation processes are insensitive to a 
tiny hyperfine splitting of molecular states. 
This is a consequence of the fact that the kinetic energy of 
colliding particles is much larger than the 
hyperfine level splitting. For the nuclear spin conversion
in molecules the situation can be  different if 
the conversion is induced by intramolecular state mixing
and collisional interruption of this mixing
(we will refer to this process as {\it quantum relaxation}
\cite{Chap96PA}) which is rather sensitive to the ortho-para level splitting. 

The CH$_3$F molecules exist in the form of two nuclear spin isomers 
\cite{Landau81}. The total spin of the three hydrogen nuclei
in the molecule can have the magnitude either $I=3/2$ (ortho isomers),
or $I=1/2$ (para isomers). Angular momentum
projections ($K$) on the molecular symmetry axis divisible by 3
are allowed only for ortho  isomers. For para isomers all other $K$ 
values are allowed.
Consequently, the quantum states of the CH$_3$F molecule are divided
into two subspaces which are schematically shown in Fig.~1. 
Qualitatively, CH$_3$F spin conversion by quantum relaxation
can be described as follows. Suppose that at the beginning of the process 
a test molecule is placed into the ortho subspace. 
Due to collisions with surrounding gas particles, which cannot 
change the spin state of the molecule, 
the test molecule starts to perform fast migration along rotational 
states inside the ortho subspace. This running up and down along the
ladder of the ortho states  continues 
until the molecule jumps into the state $m$ which is mixed
by an {\it intramolecular} perturbation $\hat V$ with the energetically
close para state $n$. During the free flight after this collision, 
para state $n$ will be admixed with the ortho state $m$. 
Consequently, the next collision can transfer the molecule
into other para states and thus localize it inside the para subspace.
More details on quantum relaxation of nuclear spin isomers can be
found in Refs.~\cite{Curl67JCP,Chap90JETP,Chap91PRA}. 

Quantum relaxation of nuclear spin isomers of molecules can be
quantitatively described in the framework of the density matrix formalism. 
The result of this description  is the following \cite{Chap91PRA}.  
One has to divide the molecular Hamiltonian into two parts
\begin{equation}
              \hat H = \hat H_0 + \hat V,
\label{H}
\end{equation}
where the main part of the Hamiltonian, $\hat H_0$, has pure ortho 
and para states as the eigenstates, and $\hat V$ is a small 
perturbation which mixes the ortho and para states. 
In the Ref.~\cite{Chap91PRA} the hyperfine contribution to the
term $\hat H_0$ was neglected and it was assumed that the molecule 
is not subjected to an external field. 

Suppose that at the instant $t=0$ a nonequilibrium 
concentration of, e.g., ortho molecules,
$\delta\rho_o(0)$, was created. the solution of the problem gives 
an exponential isomer conversion:
     $\delta\rho_o(t) = \delta\rho_o(0)e^{-\gamma t}$,
with the rate 
\begin{equation}
     \gamma = \sum_{\alpha\in o,\alpha'\in p}
         \frac{2 \Gamma \mid V_{\alpha\alpha'}\mid^2}
         {\Gamma^2 + \omega_{\alpha\alpha'}^2}
         \bigl(W_{\alpha} + W_{\alpha'}\bigr)\,.
\label{gamma}
\end{equation}
Here $\Gamma$ is the collisional decay rate of the 
off-diagonal density matrix elements, 
\begin{equation}
      (\partial\rho_{\alpha\alpha'}/\partial t)_{coll} =
        -\Gamma\rho_{\alpha\alpha'};\ \   
       \alpha \in ortho,\  \alpha' \in para,
\label{s1}
\end{equation}
assumed to be the same for all ortho-para level pairs;
$\omega_{\alpha\alpha'}$ is the gap between the states
$|\alpha>$ and $|\alpha'>$; $W_{\alpha}$ and $W_{\alpha'}$ are the Boltzmann
factors. Solution (\ref{gamma}) was obtained under the assumption
that collisions do not transfer molecules directly from ortho to para
states, i.e., the cross-section $\sigma(ortho|para)=0$. 
The validity of this assumption has been discussed in detail 
in the review \cite{Chap99ARPC}.

At low gas pressure, when $\Gamma\ll\omega$, the conversion
rate  is proportional to $\Gamma$, thus being {\it proportional} to the
gas pressure. In this limiting case, the pressure 
dependence of the conversion rate is quite similar to ordinary gaseous 
relaxation which is linear in pressure too. On the other hand,
in this pressure limit the conversion rate is sensitive to the
ortho-para level splitting (note $\omega$ in the denominator of
(\ref{gamma})) thus having a distinctive signature,
unusual for gaseous relaxation processes. This strong dependence
on $\omega$ is at the heart of the level-crossing effect at the
spin isomer conversion.

\section{Stark level splitting}

Calculation of the ortho-para state mixing is rather complicated,
especially if one has to account for both the Stark and
hyperfine level splitting. In this case special care should be taken
about proper symmetrization of states and their transformation
under these perturbations. Let us consider first the quantum states 
of the free molecule and ignore the hyperfine perturbation.
Spin-rotational states in the ground electronic and vibrational 
state of CH$_3$F can be constructed as follows \cite{Townes55,Bunker79}. 
To determine the molecular position in space, one introduces
a system of coordinates ($x$$y$$z$) fixed in the molecule 
(Fig.~2). By combining rotational and nuclear spin states,
one introduces the states $|\beta>$ which are invariant under cyclic 
permutation of the three hydrogen nuclei
\begin{equation}
                |\beta>\ \equiv\ |J,K,M>|I,\sigma,K>;\ \ 
                |\overline\beta>\ \equiv\ |J,-K,M>|I,\sigma,-K>;\ \  K\geq 0,
\label{b}
\end{equation}
where $|J,K,M>$ are the quantum states of symmetric top characterized 
by the angular momentum $(J)$, its projection $(K)$ on the $z$ axis
of the molecular system of coordinates and the angular momentum 
projection $(M)$ on the quantization $Z$ axis of the laboratory system of
coordinates. $I$ and $\sigma$ are the total spin of the three
hydrogen nuclei and its projection on the $Z$ axis, respectively.
The explicit expressions for the spin states $|I,\sigma,K>$ 
are given in \cite{Townes55}.
These expressions specify the allowed $K$-quantum numbers for the
ortho and para spin isomers as was explained in Section~II.

Permutation of any two hydrogen nuclei in CH$_3$F inverts the $z$ axis of
the molecular system of coordinates.  Such permutations, e.g.,
the permutation $P_{23}$, acts 
on the state $|\beta>$ as $P_{23}|\beta> = (-1)^{J}|\overline\beta>$.
Because the states $|\beta>$  are invariant under cyclic permutation
of the three hydrogen nuclei, a similar result is valid for the other
two permutations of hydrogen pairs: $P_{12}$ and $P_{31}$.

States $|\beta>$ and $|\overline\beta>$ generate a two-dimensional
representation of the molecular symmetry  group. 
Spin-rotation states, which generate one-dimensional representations
are
\begin{equation}
             |\beta,\kappa>\ \equiv\ \frac{1}{\sqrt{2}}
             \left[(-1)^\kappa +  P_{23}\right]|\beta>;\ \ \ 
             \kappa = 0,1.
\label{bk}
\end{equation}
From this definition it is easy to conclude that the symmetry of states
 $|\beta,\kappa>$ is determined by the rule:
$P_{23}|\beta,\kappa> = (-1)^\kappa |\beta,\kappa>$  and
by similar relations for the permutations of the other two pairs of
protons.

Further, one has to take into account the molecular inversion
states. Let us designate  the antisymmetric and symmetric inversion 
states of the molecule as $|s=0>$ and $|s=1>$, respectively.
Permutation of two protons, e.g., $P_{23}$, acts on the inversion states as
\cite{Bunker79}
\begin{equation}
             P_{23}|s=0>\ =\ -|s=0>;\ \ P_{23}|s=1>\ =\ |s=1>.
\label{P23}
\end{equation} 

The total quantum states of CH$_3$F have to be
antisymmetric under the permutation of any two hydrogen
nuclei because protons are fermions. Consequently, 
the only allowed molecular states are
$|\beta,\kappa=s>|s>$.

The last step is to account for the spin states of the 
fluorine and carbon ($^{13}$C) nuclei, both having spin 1/2. 
Finally, the molecular states are
\begin{equation}
               |\alpha_s>\ \equiv\ |\beta,\kappa=s>|s> |\sigma^F> |\sigma^C>,
\label{a}               
\end{equation}
where $\sigma^F$ and $\sigma^C$ denote, respectively, the projections
of the fluorine and carbon nuclear spins on the laboratory 
quantization $Z$ axis. The states (\ref{a}) of the free molecule will be 
denoted as the $\alpha$-basis. For rigid symmetric tops, like CH$_3$F
is, the states $|\alpha_s>$ are degenerate in the quantum numbers
$M$, $\sigma$, $\sigma^F$, $\sigma^C$ and $s$.

An external electric field due to the Stark perturbation, 
$\hat V_{St}=-\hat{\bf d}{\bf\cal E}$, mixes the states having 
$|\Delta s|=1$ and $|\Delta J|\leq 1$. In the following, we
will consider relatively weak electric fields which produce
Stark splitting much smaller than the $J$-splitting. 
Consequently, the Stark mixing of states having $|\Delta J|= 1$ 
can be neglected and one has to solve the Schr\"{o}dinger equation for the
doubly degenerate states $|\alpha_0>;\ |\alpha_1>$. In a standard 
way \cite{Landau81}, one can find the energy of the new states,
\begin{equation}
             E(\beta,\xi) = E_{free}(J,K) + 
             (-1)^\xi \frac{K |M|}{J(J+1)} d{\cal E}; \ \ \xi=0,1,
\label{Ebxi}
\end{equation}
where $E_{free}(J,K)$ is the energy of a free molecule; 
$d$ is the so-called permanent electric dipole moment
of the molecule; ${\cal E}$ is the electric field strength; 
$\xi$ is the quantum number of the new molecular states which are
\begin{equation}
     |\beta,\xi>=\frac{1}{2}\left[(-1)^\xi\frac{M}{|M|}|s=0>(1+P_{23})+
                 |s=1>(1-P_{23})\right]|\beta>.
\label{bxi}                 
\end{equation} 
The magnitude of the electric dipole moment, $d$, is determined by the
molecular spatial and electronic structure \cite{Townes55}

Similar to the case of a free molecule, one has to take into
account the spin states of fluorine  $|\sigma^F>$ and
carbon $|\sigma^C>$ nuclei which remain intact by Stark 
perturbation. Finally, the states of the CH$_3$F
molecule subjected to an external electric field are
\begin{equation}
             |\mu>\ \equiv\ |\beta,\xi>|\sigma^F>|\sigma^C>.
\label{mu}             
\end{equation}
The manifold of these states will be denoted as the $\mu$-basis.
The Stark effect partially lifts the degeneracy of the $\alpha$-states.
But the $\mu$-states remain degenerate in sign of $M$, 
and in quantum numbers  $\sigma$, $\sigma^F$ and $\sigma^C$.

\section{Hyperfine level splitting}

In accordance with the general rules, the molecular 
Hamiltonian, $\hat H$, which contains now the 
hyperfine and Stark perturbations, can be expressed
in the $\mu$-representation as
\begin{equation}
             \hat H = \sum_{\mu_1}\sum_{\mu_2}
             |\mu_2><\mu_2|\hat H|\mu_1><\mu_1|
            \ \equiv\ \hat H(o) + \hat H(p) + \hat V\ ,
\label{H1}
\end{equation}
where the sums run over the complete set of states. We recall that 
the operator $\sum|\mu><\mu|$ is the identity operator if the 
summation includes all states of the complete set.
Eq.(\ref{H1}) shows how the molecular Hamiltonian can be split  
into three operators:  $\hat H(o)$ and $\hat H(p)$  which  have 
only diagonal matrix elements in the ortho and para subspaces,
respectively, and operator $\hat V$, which has only 
matrix elements off-diagonal in the ortho and para quantum numbers.
 
When considering the combined level splitting by the three  
perturbations: Stark effect, hyperfine spin-rotation and spin-spin 
perturbations it is helpful to start with the Stark effect.
Taking into account the Stark effect is not difficult because Stark
perturbation does not mix the ortho and para states of a free molecule. 
In fact, transformation from the $\alpha$-basis of a free molecule to
the $\mu$-basis gives the level splitting by the Stark effect.

\subsection{Spin-rotation interaction}

The next perturbation to be considered is the spin-rotation coupling 
which  is due to the interaction of nuclear spins with the 
intramolecular magnetic field induced by molecular rotation.
The spin-rotation interaction Hamiltonian is given by the operator
\cite{Townes55,Wofsy71JCP}
\begin{equation}
    \hbar \hat H_{SR}\ =\
     - \sum_k\hat{\bf I}^{(k)}\bullet{\bf C}^{(k)}
     \bullet\hat{\bf J}\ \equiv\ \hbar\sum_k\hat H^{(k)}_{SR}\ ,
\label{HSR}     
\end{equation}
where $\hat{\bf I}^{(k)}$ and {\bf C}$^{(k)}$ are, respectively, 
the spin operator and  the spin-rotation tensor for the $k$-th
nucleus; $\hat{\bf J}$ is the molecular angular momentum operator 
and $k$ denotes all nuclei in the molecule. The magnitude of the
spin-rotation tensor {\bf C} depends on the molecular spatial 
structure and motion of the molecular electrons \cite{Townes55}.

Spin-rotation interaction is of the order $10-100$~kHz which
is much smaller than the energy gaps between the states having different 
$J$ and $K$ ($>10^2$~GHz), and is much smaller
than the Stark splitting of states  different in $|M|$ and $\xi$
($>10$~MHz). Consequently, only matrix elements of $\hat H_{SR}$ 
diagonal in quantum numbers $J$, $K$, $M$, $\xi$ and $I$  are
important for the calculation of the spin-rotation splitting of 
states \cite{SR}. Because the operator $\hat H_{SR}$ is a scalar, the only
non-zero matrix elements will be the elements diagonal also in the
spin projections $\sigma$, $\sigma^F$ and  $\sigma^C$.
Explicit expressions for these matrix elements can be found
in \cite{Wofsy71JCP}. For example, the diagonal matrix elements for the 
spin-rotation interaction of the fluorine nucleus are
\begin{equation}
 M\sigma^F
 \left[c_\alpha^F + \frac{K^2}{J(J+1)}(c_\beta^F-c_\alpha^F)\right],
\label{SRF}
\end{equation}
where constants $c_\alpha^F$ and $c_\beta^F$ are the
diagonal elements of the tensor {\bf C}$^F$, calculated in the
molecular system of coordinates. The spin-rotation tensor
of fluorine and hydrogen nuclei are given in \cite{Wofsy71JCP}. The 
tensor {\bf C}$^C$ for the carbon nucleus is absent in \cite{Wofsy71JCP}.
We will use for its estimation the relation:
{\bf C}$^C$={\bf C}$^F$$m^C/m^F$, where $m^C$ and $m^F$ are the
magnetic moments of $^{13}C$ and F nuclei, respectively.

One can conclude, that the spin-rotation splitting of states 
appears to be rather simple in the $\mu$-basis because
this perturbation only shifts the states but does not mix them.
As an example, the spin-rotation splitting of the state 
($J$=9,~$K$=3,~$M$=9) is shown in Fig.~3b.

\subsection{Spin-spin interaction}

Nuclear spin-spin interaction in molecules is composed of
dipole-dipole interaction of pairs of nuclei. The spin-spin interaction
Hamiltonian for the two magnetic dipoles ${\bf m}_1$ and ${\bf m}_2$
separated by the distance  $r$ has the form \cite{Landau81}
\begin{eqnarray}
     \hbar \hat H_{12} &\ =\ & 
     P_{12}\hat{\bf I}^{(1)}\hat{\bf I}^{(2)}{\ {}^\bullet_\bullet\ }
     {\bf T}^{(12)}\ ; \nonumber \\
     T_{ij}^{(12)}     &\ =\ &
     \delta _{ij}-3n_in_j\ ;\phantom{TT}P_{12}=
     m_1m_2/r^3I^{(1)}I^{(2)}\ ,
\label{H12}     
\end{eqnarray}
where $\hat{\bf I}^{(1)}$ and $\hat{\bf I}^{(2)}$ are the spin operators
of the particles 1 and 2, respectively; {\bf n} is
the unit vector directed along {\bf r}; $i$ and $j$ are
the Cartesian indices.

The total spin-spin interaction in $^{13}$CH$_3$F  ($\hat H_{SS}$)
consists of the interactions between the three hydrogen nuclei
($\hat H_{HH}$), hydrogen - fluorine nuclei  ($\hat H_{HF}$),
hydrogen - carbon nuclei ($\hat H_{HC}$), and 
fluorine - carbon nuclei ($\hat H_{FC}$). Consequently,
the total spin-spin Hamiltonian in $^{13}$CH$_3$F is
\begin{equation}
                \hat H_{SS}\ =\ \hat H_{HH} + \hat H_{HF} + 
                               \hat H_{HC} + \hat H_{FC}.   
\label{HSS}
\end{equation}
Explicit expressions for all these terms can be written on the basis
of Eq.(\ref{H12}) for one pair of nuclei. For example, for the
spin-spin interaction between the three hydrogen and fluorine
nuclei one has
\begin{equation}
              \hbar \hat H_{HF}\ =\ P_{HF}\sum_{n}\hat{\bf I}^{(n)}
          \hat{\bf I}^{F}\ {}^\bullet_\bullet\ {\bf T}^{nF}\ ;\ \ n=1,2,3\ .
\label{HF}          
\end{equation}
Here the sum runs over all hydrogen nuclei in the molecule. 
Complete definition of the spin-spin interaction requires knowledge
of the dimensional factors $P$ (see Eq.(\ref{H12})) and the spatial structure
of the molecule. These data can be found in Refs.\cite{Chap91PRA,Egawa87JMStr}.

Without hyperfine interactions,  quantum states of the molecule subjected
to an external electric field form the $\mu$-basis. The intramolecular spin-spin
perturbation $\hat H_{SS}$ has non-zero matrix elements
diagonal and off-diagonal in quantum numbers $J$, $K$, $M$,
$\sigma$, $\sigma^F$, and $\sigma^C$. 
The $J, K$-splitting of states is much larger than the spin-spin
splitting which is on the order of $10-100$~kHz. Therefore, one can 
neglect the matrix elements of $\hat H_{SS}$ off-diagonal in $J$ and $K$.

Under our conditions the Stark splitting of states different
in $|M|$ is much larger than the spin-spin interaction.
Thus, only the matrix elements of $\hat H_{SS}$  diagonal 
in $M$ are important if $|M|>1$ because of the 
selection rule $|\Delta M|\leq 2$ for the spin-spin
interaction. Off-diagonal in $M$ matrix elements of $\hat H_{SS}$
are important only for the states having $|M|=1$. This particular
case will be considered elsewhere. One can conclude, that the 
spin-spin perturbation of $\mu$-states
is determined by the matrix elements of $\hat H_{SS}$ diagonal in
quantum numbers $J$, $K$, $M$, $\xi$, $I$ and in the values of 
the sum $\sigma + \sigma^F +\sigma^C$. 

To summarize, the hyperfine structure of molecular states in our case
is determined by the hyperfine Hamiltonian $\hat H_{SR}+\hat H_{SS}$
spanned by the states different only in the projections of nuclear
spins $\sigma$, $\sigma^F$, and $\sigma^C$. There are $4\times2\times2=16$ such
states for ortho molecules:
\begin{equation}
     \mid\beta_{ortho},\xi>\mid\sigma^F>\mid\sigma^C>,
\label{bortho}
\end{equation}
and $2\times2\times2=8$ states for para molecules:
\begin{equation}
     \mid\beta_{para}\xi>\mid\sigma^F>\mid\sigma^C>.
\label{bpara}
\end{equation}

Matrix elements of $\hat H_{SR}$ are give by Eq.~(\ref{SRF}).
Calculation of the matrix elements of all terms of $\hat H_{SS}$ 
can be done in a way similar to that explained in the Appendix. 
After all necessary matrix elements of $\hat H_{SR}$ and $\hat H_{SS}$
are determined one can find the eigenstates and eigenvalues
of the operator $\hat H_{SR} + \hat H_{SS}$. This will give 
the hyperfine level splitting
of the  molecular states under joint action of the spin-spin and 
spin-rotation interactions. In the case of ortho states one
has to diagonalize the 16x16 matrix. For the para states one has
to diagonalize the 8x8 matrix. Ortho and para quantum states of
the molecule which take into account the Stark effect and both types of 
hyperfine interactions will be designated as the $\tau$-basis. As 
an example, splitting of the ortho state $J$=9, $K$=3, $M$=9
by the spin-spin and spin-rotation interactions is shown in Fig.~3c. 
These calculations were done numerically. One can see from these data
that the hyperfine interaction lifts completely degeneracy of the
nuclear spin states. 

\section{Conversion spectra}

\subsection{Rotational structure in the conversion spectra}

For the calculation of the conversion spectra
it is important to know the positions of ortho and para states in 
$^{13}$CH$_3$F. The nuclear spin conversion in $^{13}$CH$_3$F is dominated by
mixing of the two level pairs: ($J$=9,~$K$=3)--($J'$=11,~$K'$=1) 
and (20,3)--(21,1) \cite{Chap91PRA,Chap93CPL}.
Of these two level pairs the former has the smaller energy gap (130~MHz) and
contributes nearly 65\% to the conversion rate of a free molecule. The 
states (11,1)--(9,3) are mixed
by the spin-spin interaction only \cite{Chap91PRA}. The 
spin-rotation interaction  does not mix this pair of states
because of the selection rule for the spin-rotation interaction 
$|\Delta J|\leq 1$ \cite{Guskov95JETP}.

To describe the rotational structure in the conversion spectra
\cite{Nagels95CPL} one has to include
the Stark perturbation $\hat V_{St}=-\hat{\bf d}{\bf\cal E}$ 
into the $\hat H_0$ term of the splitting (1). As a result,
an equation similar to Eq.(\ref{gamma}) is obtained 
but in the $\mu$-basis which accounts for 
the Stark effect. Note that the modeling of the off-diagonal 
elements of the collision integral (\ref{s1}) will also have 
the same form in the $\mu$-basis. Further, one needs to know 
the matrix elements of the spin-spin perturbation
which mixes  ortho and para states of $^{13}$CH$_3$F. From the
total spin-spin perturbation (\ref{HSS}) only the part, 
$\hat V = \hat H_{HH} + \hat H_{HF} + \hat H_{HC}$,
produces the ortho-para state mixing. The matrix elements 
$V_{\mu'\mu}$ in the $\mu$-basis can be calculated using
the matrix elements of $\hat V$ in the $\alpha$-basis \cite{Chap91PRA}
and the relation between these two bases given in Section~III. 

An overview of the rotational structure in conversion spectra of 
$^{13}$CH$_3$F is shown in Fig.~4. The spectrum was calculated using 
$\Gamma=1.75\cdot10^7$~s$^{-1}$ which corresponds to the gas
pressure 0.1~Torr. (Here and below we use the decoherence rate
$1.75\cdot10^8$~s$^{-1}$/Torr \cite{Nagels96PRL}). 
For such a value of $\Gamma$ the hyperfine structure 
of molecular states is not important because the line broadening 
is much larger than the hyperfine splitting. At lower pressures
the hyperfine level splitting does play an important role,
as is shown in the next Section.

\subsection{Hyperfine spectra}

To calculate the hyperfine spectra of spin conversion
one has to  include both the Stark terms
and the hyperfine terms into the operators $\hat H(o)$
and $\hat H(p)$ (see Eq.(\ref{H1})). This will result in 
transformation of the $\mu$-basis of molecular states to the 
$\tau$-basis. Matrix elements of the perturbation $\hat V$
between ortho and para states should be calculated in the $\tau$-basis
also. The calculations were done by using the matrix elements of 
$\hat V$ in the $\alpha$-basis from \cite{Chap91PRA} and the matrix
which transforms the $\alpha$-basis to the $\tau$-basis. 

The new expression for the conversion rate obtained in this way, 
which is exactly Eq.(\ref{gamma}) with $\tau$ indices instead of
$\alpha$ indices, needs further modification. The point is that  
hyperfine structure in the spin conversion spectra
can be revealed only at rather low pressures. The gas pressure
should be low enough to make the decoherence rate 
$\Gamma$ smaller than the hyperfine splitting.
The latter is estimated to be of the order of $V$. Consequently, 
the condition $\Gamma\lesssim V$ should be fulfilled. 

As was shown in \cite{Chap96PA}, calculation of the spin
conversion rate in first order perturbation theory is not
valid if $\Gamma\lesssim V$. This  is a consequence of the 
level population saturation  by intramolecular ortho-para 
states mixing. The saturation effect can be accounted for 
by using the new expression for the conversion rate
\begin{equation}
        \gamma = \sum_{\tau'\in p, \tau\in o}
        \frac{2 \Gamma_{\tau'\tau}|V_{\tau'\tau}|^2}
        {\Gamma^2_{\tau'\tau} + \omega^2_{\tau'\tau} +
        4\frac{\Gamma_{\tau'\tau}}{\nu}|V_{\tau'\tau}|^2}
        \left(W_{\tau'} + W_{\tau}\right),
\label{gsat}        
\end{equation}
where $\nu$ is the rotational relaxation rate which was assumed
equal for ortho and para isomers. Expression (\ref{gsat}) 
is a straightforward generalization of the result \cite{Chap96PA}, where 
the theory was developed for the mixing of only one pair of
ortho and para states.

As is clear from Eq.~(\ref{gsat}), the saturation effect 
is most important for ``strong resonances'', which
have large values of the mixing matrix elements $V_{\tau'\tau}$.
As an example of the saturation effect at work, we show the
hyperfine structure of the strongest peak in the spectrum of
Fig.~4 which results from the crossing of magnetic sublevels $M'$=11 
and $M$=9. This hyperfine spectrum is given in Fig.~5. Calculation 
of the spectrum was performed for the gas pressure 0.1~mTorr, 
and the ratio $\Gamma/\nu = 10$. From the spectrum of Fig.~5
one concludes that the saturation effect can produce substantial
broadening of the spectral lines. Consequently, the hyperfine structure
in Fig.~5 remains unresolved. Decreasing the gas pressure does not result 
in a decrease in the width of the lines. 

Nevertheless, well-resolved hyperfine spectra of spin conversion
can be obtained if one chooses ``weak resonances'' having
small magnitude of $V_{\tau'\tau}$. As an example, the hyperfine
structure produced by the ``weak crossing'' of the $M'$=7 and $M$=9
states is shown in Fig.~6. One can see that in this case the hyperfine
structure is well resolved. Decreasing the gas pressure makes the 
lines even more narrow.

It is interesting to understand the accuracy of the simplified approach
to the conversion spectra in which hyperfine splitting of molecular states
is not taken into account. In Fig.~7 we show the pressure dependence of 
the amplitude of the ``strong resonance'' (9,3,9)--(11,1,11) calculated
using the simplified and the present model. One can see that a 10\%
difference in the two amplitudes already appears at 5~mTorr. At
the pressure 1~mTorr, the present model gives one-third the value
of the old model.
 
\section{Conclusions}

A theoretical model of hyperfine spectra of nuclear
spin conversion has been developed. Calculations 
have been performed for the conversion of  spin isomers
of $^{13}$CH$_3$F molecules subjected to an external electric 
field. The spin-rotation and the spin-spin intramolecular 
interactions were taken into account in the calculations.

It has been shown that the hyperfine structure of spin
conversion spectra can be substantially distorted by the saturation 
effect. In the case of ``strong resonances'' this effect
hides the hyperfine structure of the spectra completely.
It had been proposed to observe the hyperfine spectra 
by using ``weak resonances'' which have small ortho-para matrix elements. 
Such weak resonances have been found for the $^{13}$CH$_3$F molecules 
and hyperfine spectra for them have been calculated.

Calculations performed in the present paper have revealed 
the conditions at which experimental observation
of the hyperfine spectra of nuclear spin conversion can be performed.
These conditions include the choice of suitable level crossings
and gas pressure. The pressure at which the hyperfine structure of 
the spectra is resolved appeared to be rather low. This implies 
strong limitations to the experimental setup, viz., molecular 
collisions with the walls should contribute to the
decoherence rate $\Gamma$ not much more than collisions in the bulk. 
The latter should be on the order of $10^4$~s$^{-1}$ only.

The theoretical model proposed in this paper is rather general. 
It can be applied to molecules having various symmetries without 
changing its essence. Another possible extension of the model can 
be related to the use of the Zeeman effect for the level splitting 
instead of the Stark effect considered in the paper. 

\section{Acknowledgment} 
 
This work  was made possible by financial support from the Russian 
Foundation for Basic Research (RFBR), grant No. 98--03--33124a and from 
the Region Nord Pas de Calais, France.

\section{Appendix}

As an example, we give here the calculation of the matrix elements 
of $\hat H_{HF}$. First, let us express the $\hat H_{HF}$ through 
the spherical tensors \cite{Landau81}:
\begin{equation}
     \hat H_{HF}\ \ \equiv\ \sum_n \hat H^{(n)}_{HF}\ =\ 
     P_{HF}\sum_n\sum_{q_1,q_2,q}
     (-1)^q<1,q_1,1,q_2,|2,-q>
     \hat I^{(n)}_{1,q_1}\hat I^F_{1,q_2}T^{nF}_{2q}.
\label{HHF1}
\end{equation}
Here $\hat H^{(n)}_{HF}$ is
the spin-spin operator of the interaction between fluorine and
the $n$-th hydrogen nuclei; $n=1,2,3$ denotes the hydrogen nuclei;
$<\dots|\dots>$ stands for the Clebsch-Gordan coefficient. 

Calculations of the matrix elements of $\hat H_{HF}$ can be 
substantially simplified if one takes into account that the
matrix elements $<\mu_1|\hat H^{(n)}_{HF}|\mu_2>$ are equal 
for all $n$. This can be proven by applying to these matrix 
elements the cyclic permutation, $P_{123}$, of the three equivalent 
protons. Taking this simplification into account one has
\begin{equation}
<\mu_1|\hat H_{HF}|\mu_2>\ =\ 3<\mu_1|\hat H^{(1)}_{HF}|\mu_2>.
\label{3}
\end{equation}

The matrix elements of the operator $\hat H_{HF}$ which contribute to
the spin-spin level splitting (see Section~IV) are given by the expression
\begin{eqnarray}
     <\mu_2|H_{HF}|\mu_1> &\ =&\ 3P_{HF}\sum_q<1,q,1,-q|2,0>
     <I,\sigma_2,K|\hat I^{(1)}_{1,-q}|I,\sigma_1,K>\times  \nonumber \\
      && <\sigma^F_2|\hat I^F_{1,q}|\sigma^F_1> 
     <J,K,M|T^{1F}_{2,0}|J,K,M> \delta_{\sigma^C_2,\sigma^C_1}.  
\label{HF2}
\end{eqnarray}
The matrix elements in Eq.(\ref{HF2}) can be evaluated using the following 
relations. The matrix elements of the spatial tensor $T^{1F}_{2,0}$ are 
given in Ref.~\cite{Landau81}. In our particular case one has
\begin{equation}
 <J,K,M|T^{1F}_{2,0}|J,K,M>\ =\ <J,K,2,0|J,K><J,M,2,0|J,M>{\cal T}^{1F}_{2,0},                     
\label{T20}
\end{equation}
where ${\cal T}^{1F}_{2,0}$ is the spherical tensor component of
${\bf T}^{1F}$ calculated in the molecular frame. Finally,
the matrix elements of the spin operators $\hat I^{(1)}_{1,-q}$
and $\hat I^F_{1,q}$ can be evaluated using the
Wigner-Eckart theorem which states for the spherical tensor
of rank $\kappa$  \cite{Landau81}:
\begin{equation}
<J',M'|f_{\kappa,q}|J,M>\ =\ {\it i}^\kappa(-1)^{J_{max}-M'}
                 \left(\begin{array}{rcc} J' & \kappa & J \\
                                   -M' & q & M\\\end{array}\right)
                                    <J'||f_\kappa||J>,
\label{WE}
\end{equation}
where (:\ :\ :) stands for a 3j-symbol and $<J'||f_\kappa||J>$ is the 
reduced matrix element. Eqs.(\ref{HF2})-(\ref{WE}) allow us to perform 
the calculations of the matrix elements of $\hat H_{HF}$.

\newpage
\begin{figure}[htb]
\centerline{\psfig
{figure=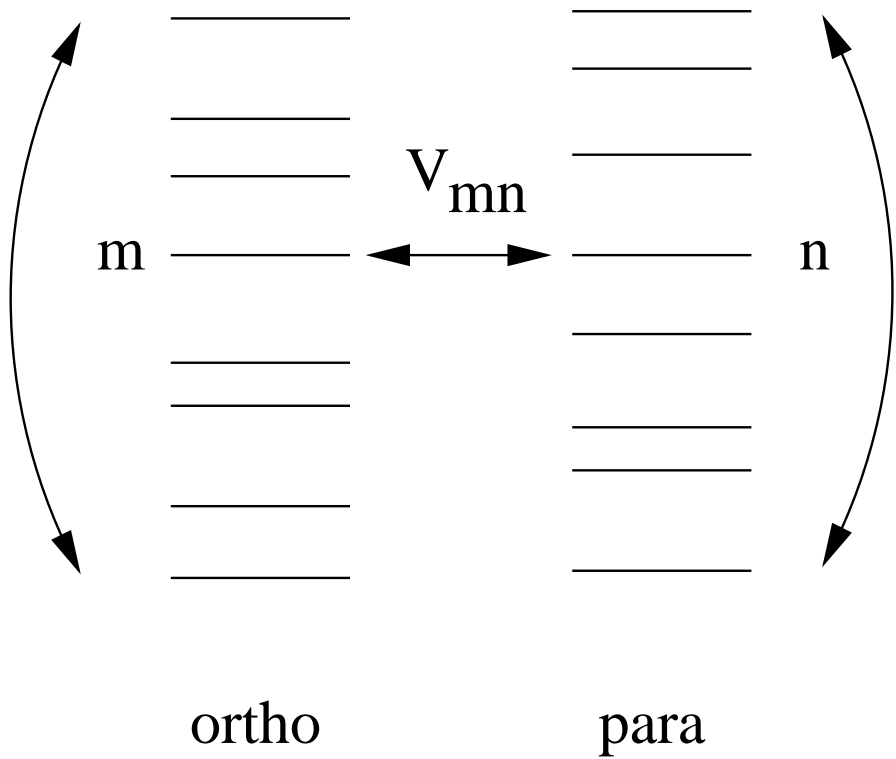,height=10cm}}
\vspace{3cm}
\caption{\sl \label{fig1}
Schematic of the ortho and para states of the CH$_3$F molecule.
It is assumed that only one pair of ortho and para states ($m-n$) is
mixed by an intramolecular perturbation $\hat V$. The bent
lines indicate transitions induced by collisions with
gas particles. These collisions do not produce direct transitions 
between molecular ortho and para states.}
\end{figure}

\newpage
\begin{figure}[htb]
\centerline{\psfig
{figure=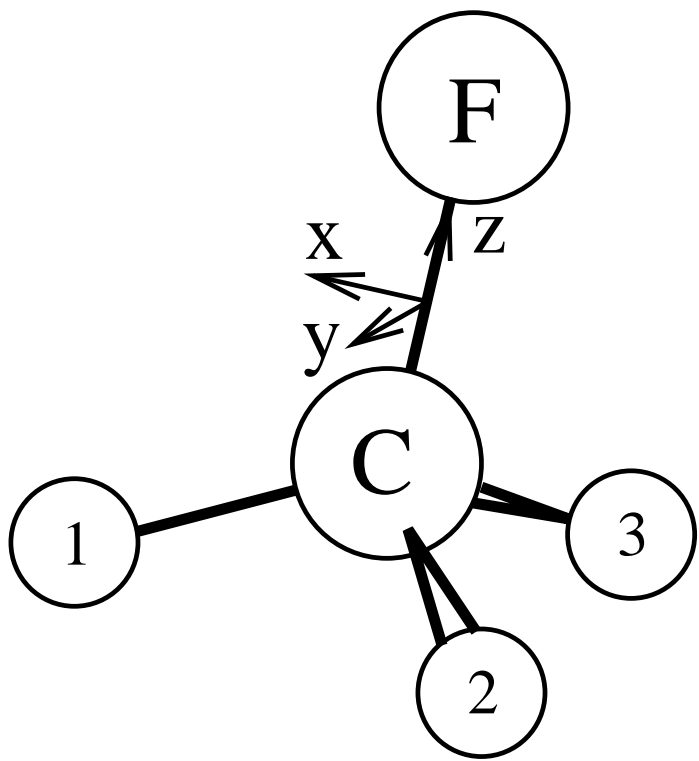,height=10cm}}
\vspace{3cm}
\caption{\sl \label{fig2}
Orientation of the molecular system of coordinates.
It has its origin in the molecular centre of mass and is oriented
by the numbered hydrogen nuclei in such a way that the $xy$ plane 
is parallel to the hydrogen plane, the $x$ axis is directed to 
the H$^{1}$ atom and the $y$ axis is between the H$^{1}$ and H$^{2}$ 
atoms. The $z$ axis is directed along the molecular symmetry axis.}
\end{figure}

\newpage
\begin{figure}[htb]
\centerline{\psfig
{figure=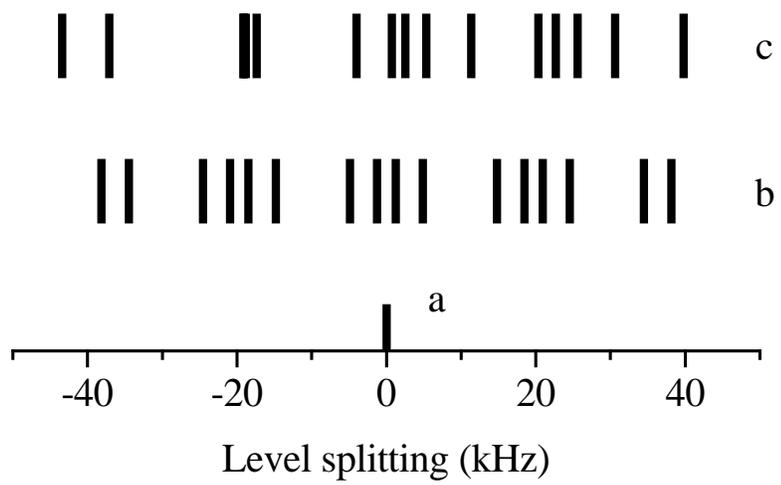,height=20cm}}
\vspace{-4cm}
\caption{\sl \label{fig3}
Splitting of the state $J$=9, $K$=3, $M$=9 of $^{13}$CH$_3$F: 
(a) no hyperfine perturbation; (b) splitting under the spin-rotation 
perturbation; (c) splitting by joint action of the spin-rotation and 
spin-spin  perturbations. Note that three components near 19~kHz are 
accidentally nearly degenerate.}
\end{figure}

\newpage
\begin{figure}[htb]
\centerline{\psfig
{figure=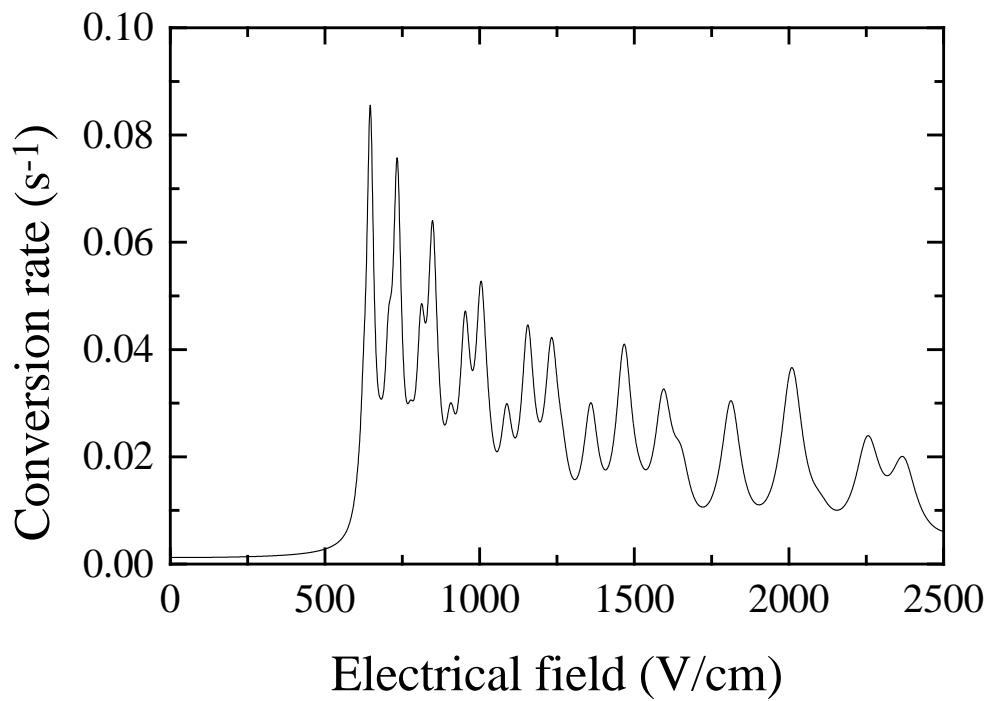,height=20cm}}
\vspace{-1cm}
\caption{\sl \label{fig4}
Overview of the level-crossing resonances in the conversion
({\it conversion spectra}) of $^{13}$CH$_3$F nuclear spin 
isomers. The decoherence rate in this example was taken equal 
$\Gamma=1.75\cdot10^7$~s$^{-1}$ (gas pressure 0.1~Torr).}
\end{figure}

\newpage
\begin{figure}[htb]
\centerline{\psfig
{figure=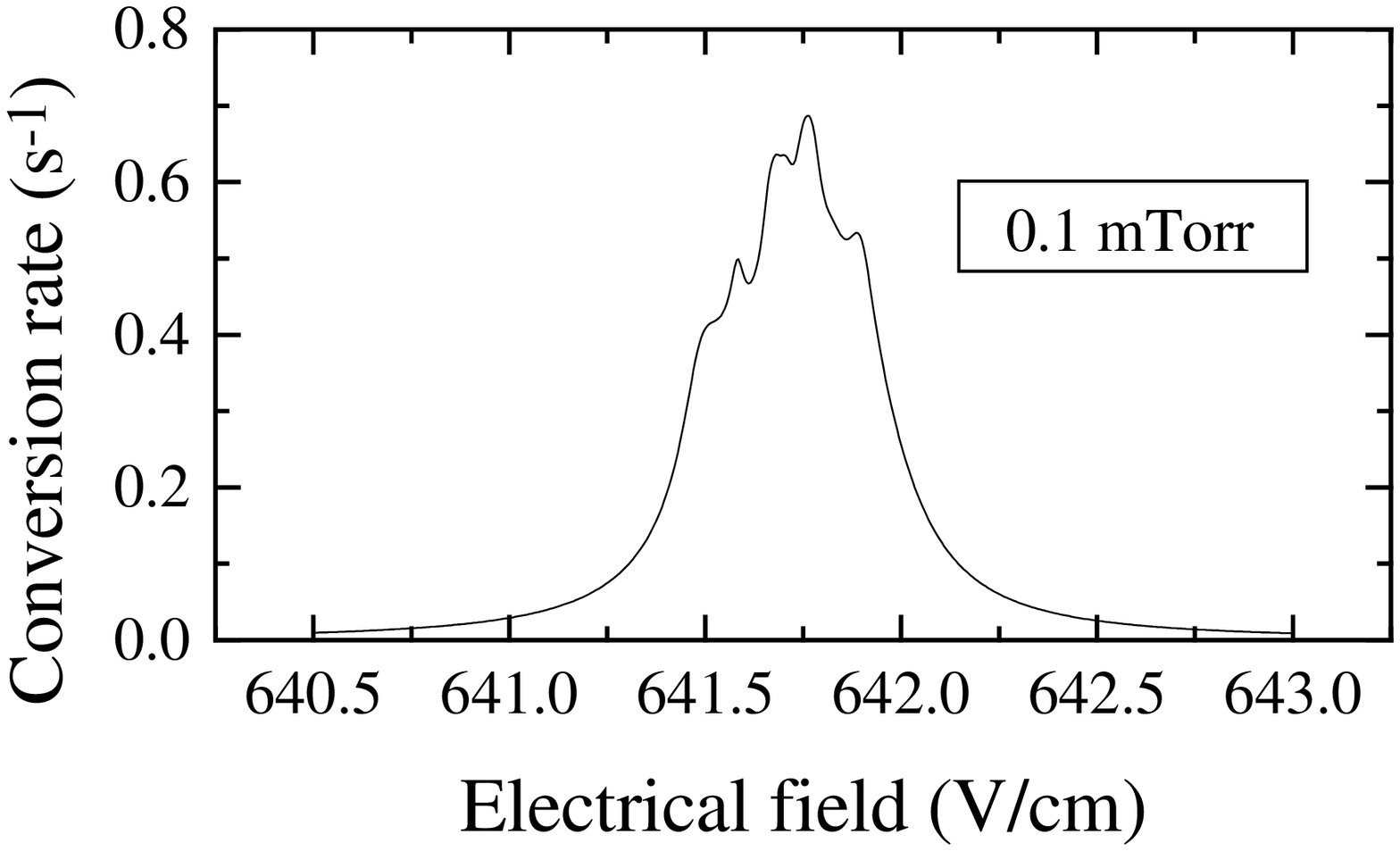,height=20cm}}
\vspace{-1cm}
\caption{\sl \label{fig5}
Hyperfine structure of the ``strong resonance'' resulting 
from the crossing of  magnetic sublevels $M'$=11 and $M$=9. 
The gas pressure is 0.1~mTorr.}
\end{figure}

\newpage
\begin{figure}[htb]
\centerline{\psfig
{figure=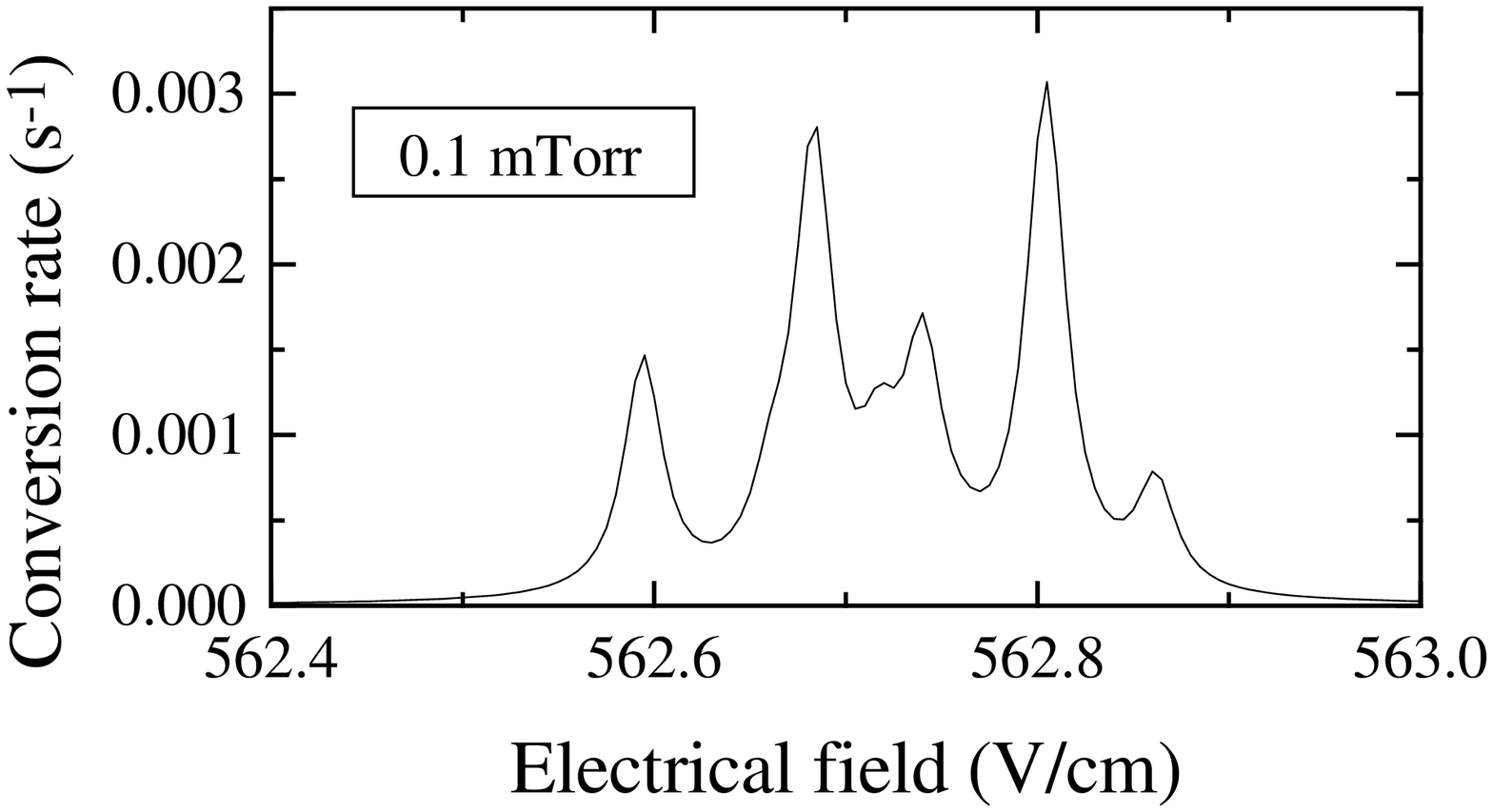,height=20cm}}
\vspace{-1cm}
\caption{\sl \label{fig6}
Hyperfine structure of the ``weak resonance'' resulting 
from the crossing of  magnetic sublevels $M'$=7 and $M$=9. 
The gas pressure is 0.1~mTorr.} 
\end{figure}

\newpage
\begin{figure}[htb]
\centerline{\psfig
{figure=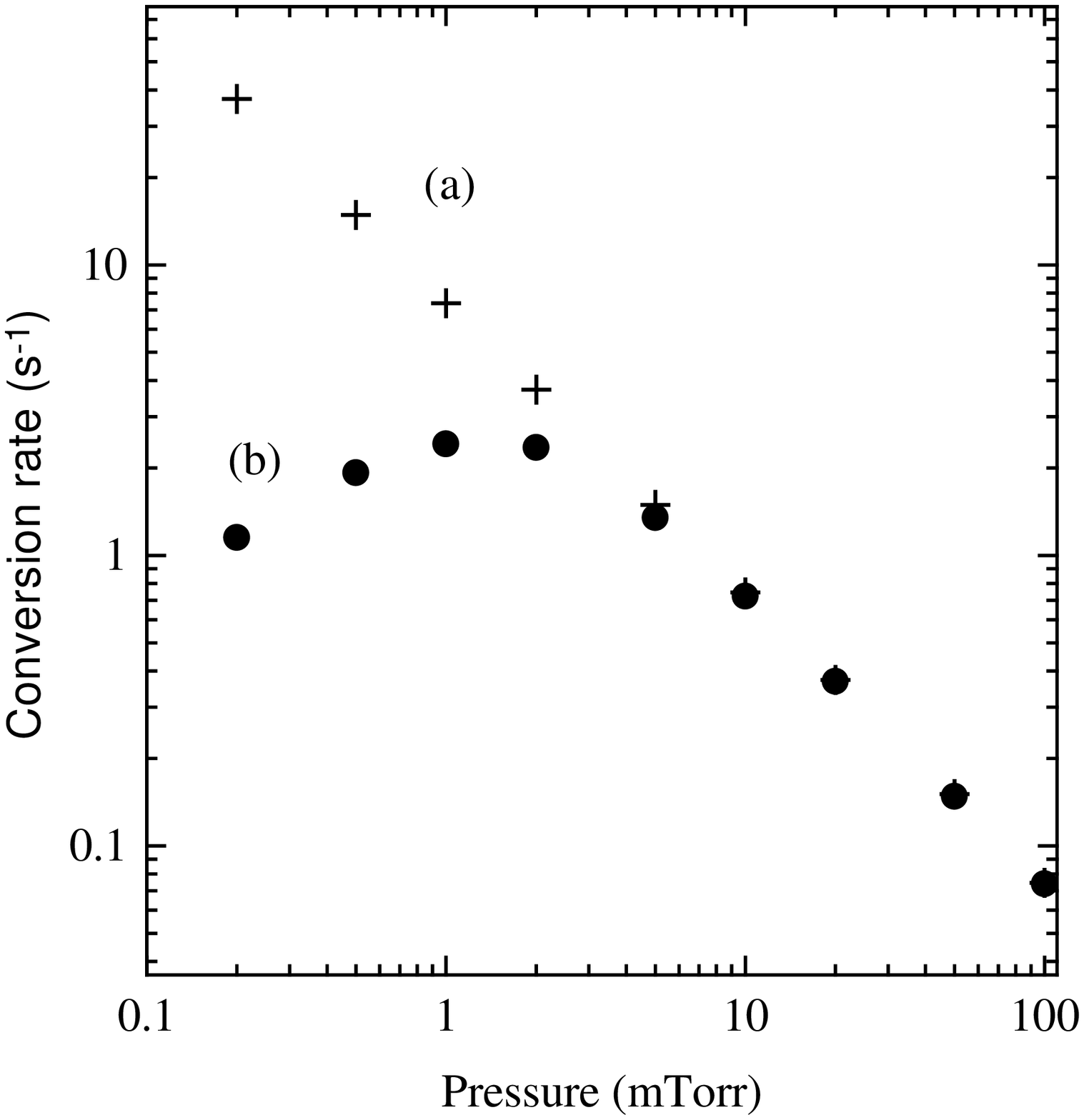,height=20cm}}
\vspace{0.5cm}
\caption{\sl \label{fig7}
Pressure dependence of the amplitude of the ``strong
resonance'' (9,3,9)--(11,1,11) (a) without and (b) with 
accounting of the hyperfine splitting of molecular states.}
\end{figure}


\begin{thebibliography}{10}
\bibitem{SR}
The conclusion concerning $M$ is the consequence of the selection rule for the
  spin-rotation interaction $|\Delta M|\leq 1$ \cite{Guskov95JETP}.
\end{thebibliography}
\end{document}